# Assessment of cloud and associated radiation fields from a GAN stochastic cloud subcolumn generator

**Dongmin Lee[1,2], Lazaros Oreopoulos[2], Nayeong Cho[2,3], and Daeho Jin[2,3]**

[1]*GESTAR-II, Morgan State University, Baltimore, MD, USA.*

[2]*Earth Sciences Division, NASA's Goddard Space Flight Center, Greenbelt, MD, USA.*

[3]*GESTAR-II, University of Maryland – Baltimore County, MD, USA.*

## Abstract

Modern Earth System Models (ESMs) operate on horizontal scales far larger than typical cloud features, requiring stochastic subcolumn generators to represent subgrid horizontal and vertical cloud variability. Traditional physically-based generators often rely on analytical cloud overlap paradigms, such as exponential-random decorrelation, which can struggle to capture the complex, anti-correlated behavior of non-contiguous cloud layers. In this study, we introduce a novel two-stage machine learning subcolumn generator for the GEOS atmospheric model, utilizing a Conditional Variational Autoencoder combined with a Generative Adversarial Network (CVAE-GAN) and a U-Net architecture. Trained on a merged CloudSat-CALIPSO height-resolved cloud optical depth dataset, the ML generator creates 56 stochastic subcolumns representing cloud occurrence and optical depth profiles. Evaluated against the established Räisänen generator, the ML approach accurately reproduces bimodal cloud overlap distributions, significantly reduces biases in grid-mean statistics, and halves the root-mean-square error in ISCCP-style cloud-top pressure and optical thickness joint histograms. The improvements brought by our deep generative models translate into more accurate offline radiative transfer calculations, reducing the global-mean shortwave top-of-atmosphere cloud radiative effect bias by a factor of three. Provided that the generator can be accelerated on CPUs, this offers a practical pathway to reduce structural errors at the cloud-radiation interface.

## 1. Introduction

Modern atmospheric GCMs solve their physics on horizontal grids of order 50-100 km, scales much larger than typical cloud features. Cloud-radiation interactions, precipitation, and aerosol activation all depend nonlinearly on cloud properties, so representing them accurately requires some accounting of the subgrid horizontal and vertical cloud variability that is not resolved by the model's prognostic cloud fields (Pincus et al., 2003).

One approach to bridge this scale gap is a stochastic cloud subcolumn generator, which produces an ensemble of cloudy subcolumns whose mean closely matches the grid-mean cloud profile while their distribution captures subgrid heterogeneity. In NASA's Goddard Earth Observing System (GEOS) global model, subcolumns generated this way feed directly into the radiative transfer scheme via the Monte Carlo Independent Column Approximation (McICA, Pincus et al., 2003), which couples the radiation calculation to a single subcolumn at each correlated-k g-point. Subcolumns, but not necessarily those seen by the radiation scheme, are also used by satellite simulators such as COSP (Bodas-Salcedo et al., 2011) when evaluating modeled clouds against observed cloud fields.

Two stochastic subcolumn generators are widely used in this context. The Subgrid Cloud Overlap Profile Sampler (SCOPS, Klein & Jakob, 1999), as implemented in COSP, generates a binary cloud mask under maximum-random overlap and assigns every cloudy cell at a given level the layer-mean optical depth, so each layer is horizontally homogeneous. The generator of Räisänen et al. (2004) instead uses the generalized (exponential-random) overlap paradigm of Hogan and Illingworth (2000), in which the weighting between maximum and random overlap of cloud occurrence decays exponentially with vertical separation distance, with a characteristic decorrelation length. Räisänen subcolumns also carry horizontal cloud optical depth (COD) variability within each layer, drawn from a beta probability density function (PDF) with vertically correlated ranks.

The two decorrelation lengths in the Räisänen generator - one for cloud occurrence overlap and one for the COD PDF rank correlation - are parameterized in GEOS as Gaussian functions of latitude and day of year, fitted to observed cloud overlap statistics (Oreopoulos et al., 2012). Recent observational evaluations have made it possible to confront these two generators directly with the cloud subgrid variability they aim to represent. Oreopoulos et al. (2022a, hereafter O22a) constructed a CloudSat-CALIPSO reference dataset of two-dimensional vertically-resolved cloud optical depth fields by merging the liquid-phase 2B-CWC-RVOD and ice-phase 2C-ICE retrievals into a single along-track product, with missing liquid retrievals filled using collocated MODIS-Aqua total optical depth as a constraint.

Treating each ~110 km along-track segment as a "scene", and conditioning each generator on the corresponding mean profiles of cloud fraction and COD, O22a evaluated the simulated subgrid variability against observations in the cloud-top pressure / cloud optical thickness (CTP-TAU) joint histogram phase space. Oreopoulos et al. (2022b, hereafter O22b) used the same merged dataset to revisit the Räisänen decorrelation-length parameterization, deriving updated Gaussian fits in latitude and day of year that yield larger peak decorrelation lengths in the tropics than the earlier parameterization of Oreopoulos et al. (2012).

Both studies converge on similar bias characteristics. SCOPS systematically underestimates optically thin clouds while overestimating clouds of moderate to high optical thickness, with an associated shortwave cloud radiative effect (CRE) bias of about 3 W $m^{-2}$ on globally averaged oceanic scenes. The Räisänen generator behaves better and yields a small global SW CRE bias, but it still underestimates the optically thinnest clouds and overestimates total cloud occurrence overlap. Even when the Räisänen overlap parameterization is re-tuned to the same observational dataset, the improvement over the older parameterization is marginal, suggesting that the residual error is due to the structural weakness of the generator rather than the values of its specific coefficients (O22b).

These biases are not merely cosmetic. Lee and Oreopoulos (2025) recently showed that, in a single AMIP4K perturbation experiment with GEOS, the global cloud feedback derived through the cloud radiative kernel framework varies by nearly a factor of three 0.4-1.1 W $m^{-2}$ $K^{-1}$ across permutations of subcolumn generator, joint-histogram interpretation, and kernel. The subcolumn generator alone contributes a substantial fraction of this spread. Reducing the residual bias of the underlying subcolumn generator is therefore a direct lever on the spread of model-derived cloud feedback estimates.

Machine learning offers a route that is in principle better suited to capturing structured subgrid variability than the analytical overlap and PDF parameterizations. Indeed, recent studies have increasingly turned to ML and high-resolution observational data, such as CloudSat, to improve the representation of subgrid cloud structures. For instance, neural networks have been employed to diagnose cloud fraction (Chen et al., 2023; Pi et al., 2026), and deep learning has been used to simulate cloud vertical overlap structures (Wu et al., 2026), demonstrating improved predictive performance over traditional empirical methods. Other efforts have shown that ML can effectively infer fine-scale convective cloud information from coarse-resolution environmental conditions (Yu et al., 2026). However, these existing ML-based schemes primarily focus on diagnosing macroscopic cloud properties such as cloud fraction or overlap. And do not directly predict subgrid cloud water content or optical depth. Alongside these developments, recent work has shown that physically informed neural networks can emulate the full RTE+RRTMGP radiation scheme with offline accuracy of a few percent across cloudy and clear-sky conditions while learning physically meaningful input-output relationships (Hafner et al., 2025), suggesting that ML is mature enough to be considered for related cloud-radiation interface problems. The subcolumn generation task is, however, distinct from radiation emulation: it is not a deterministic regression problem but a conditional generative one, in which only marginal distributions and overlap statistics of the output are observable, and any single realization is one of many that satisfy the same large-scale mean profiles. Capturing this generative structure correctly is the central modeling challenge.

In this work we develop an ML subcolumn generator for the GEOS model that takes the same large-scale grid-mean inputs as the Räisänen generator and produces 56 stochastic subcolumns of cloud occurrence and ice/liquid optical depth at the native MERRA-2 vertical grid. We train the generator against the merged CloudSat-CALIPSO reference dataset of O22a, and evaluate its performance via the joint CTP-TAU histogram as in O22a and O22b. Throughout the paper we benchmark the ML generator against the Räisänen generator with the updated O22b decorrelation lengths, which is the strongest stochastic baseline currently available for this dataset. SCOPS is not retained as a comparison because O22a already established that it is the weaker of the two stochastic generators on this exact reference, and the goal of this work is to ask whether an ML approach can improve on the best available physically motivated option.

## 2. Data

### 2.1 Reference cloud fields from CloudSat-CALIPSO

The reference vertically resolved cloud fields used to train and evaluate the generators are the merged two-dimensional cloud optical depth (COD) dataset of O22a, with the overlap statistics revisited in O22b. The dataset combines three CloudSat release-5 products: the liquid-phase 2B-CWC-RVOD product (Leinonen et al., 2016), the CALIPSO-enhanced ice-phase 2C-ICE product (Deng et al., 2015), and the 2B-CLDCLASS-LIDAR cloud mask (Sassen and Wang, 2008, 2012).

Each CloudSat ray (subcolumn) carries a vertically resolved profile of ice COD, liquid COD, and cloud thermodynamic phase, at the native 240 m × 1.7 km cell resolution of the CloudSat Cloud Profiling Radar. Because the radar-only liquid phase retrievals in 2B-CWC-RVOD fail for a non-negligible fraction of cells that are still flagged as liquid cloud in the CALIOP-informed 2B-CLDCLASS-LIDAR product, O22a apply a two-pass filling scheme that assigns liquid COD values to these missing cells using the collocated MODIS-Aqua total optical depth as a constraint and, where that is unavailable, a nearest-neighbor interpolation.

We adopt the filled field as our reference throughout. For 2007 the dataset contains approximately $5.86 \times 10^5$ oceanic 100-subcolumn scenes (~110 km along-track each). For use here we organize the dataset into fixed 56-ray scenes rather than the 100-ray scenes of O22a. The smaller segment length aligns the scene footprint with the 3-hourly Modern-Era Retrospective Analysis for Research and Applications, version 2 (MERRA-2) grid box used as input (see next subsection) and lets every scene be matched cleanly to a single MERRA-2 grid column without spatial averaging at the boundary. As shown in O22a, scene size above ~20 rays has little effect on the aggregate joint-histogram statistics we use for evaluation, so this restructuring does not compromise the comparison protocol.

Three target quantities are retained per subcolumn cell: a binary cloud mask, the in-cloud ice COD, and the in-cloud liquid COD. Effective radii are not modeled, instead, their scene-mean values from the reference dataset are applied uniformly across the subcolumns during the downstream offline radiative transfer calculations.

### 2.2 Atmospheric state from MERRA-2

Large-scale atmospheric state variables used to condition both generators are taken from NASA's MERRA-2 reanalysis (Gelaro et al., 2017). Specifically, we use the inst3_3d_asm_Nv product which provides 3-hourly instantaneous 3D assimilated fields on the 72 native model hybrid-sigma levels at 0.5° × 0.625° horizontal resolution.

For each CloudSat ray we perform a nearest-neighbor lookup in (latitude, longitude, time) to retrieve the collocated MERRA-2 column. We use the mid-layer geopotential height, pressure, temperature, specific humidity, relative humidity, ozone mixing ratio, and the zonal and meridional wind components. The CloudSat cloud profile, originally on a 240 m vertical grid, is remapped onto the MERRA-2 hybrid-sigma grid using log-pressure linear interpolation. COD is conserved in this remapping by converting to extinction per unit thickness before interpolation and multiplying back by the new layer thickness.

Scene-level scalars are also retained from CloudSat and MERRA-2: land-sea mask, latitude, longitude, surface pressure, skin temperature, solar zenith angle, surface albedo, and Julian day. We restrict the vertical domain to the lowest 39 MERRA-2 layers (approximately the surface to ~50 hPa), which contains essentially all observed cloud in the reference dataset. Above this level the scene-averaged cloud fraction is below 1% and the generators have nothing to predict.

Because boundary-layer decks and deeper cumulus respond differently to the vertical wind structure, we convert the MERRA-2 zonal and meridional wind to shear profiles by subtracting the value at the lowest model layer. The resulting profiles capture the vertical wind structure relevant to cloud organization without carrying the surface-reference mean wind itself, which is not expected to control subgrid overlap directly.

After vertical restriction, per-subcolumn variable selection (mask, ice COD, liquid COD), and rejection of scenes containing no cloud anywhere in the 39-layer column, the final dataset comprises approximately $1.8 \times 10^6$ scenes of 56 subcolumns each over 2008-2010 for training and validation, split 90/10. The year 2007 is held out entirely and used for evaluation against the Räisänen baseline, matching the evaluation period of O22a and O22b.

## 3. Methods

### 3.1 Problem framing

We frame subcolumn generation as a conditional generative problem. A scene consists of the 39-layer large-scale atmospheric column together with the 39-layer profiles of layer cloud fraction (CF), layer-mean ice COD, and layer-mean liquid COD. The generator target is a set of 56 subcolumns, each of which is a 39-layer profile of (mask, ice COD, liquid COD), such that:

1. the subcolumn-mean profile of mask, ice COD, and liquid COD matches the large-scale input to the scene, and
2. the scene-level statistics of cloud overlap, column optical depth, and the joint (CTP, TAU) distribution match the observed statistics computed from the CloudSat-CALIPSO reference of O22a.

Every method we compare, the stochastic generator of Räisänen et al. (2004), the ML generator developed here, and any ablation, receives the same 39-layer large-scale profiles and produces a set of 56 subcolumns as output. Evaluation in the cloud top pressure-cloud optical-thickness (CTP-TAU) joint histogram phase space follows O22a and O22b, with ISCCP-standard CTP and TAU bin edges throughout, but using direct cloud-top pressure, so that any improvement attributable to the ML generator is directly comparable to the SCOPS-vs-Räisänen contrast documented in those studies.

### 3.2 Räisänen generator

As the physically based baseline we use the stochastic subcolumn generator of Räisänen et al. (2004), configured identically to O22a and O22b. Given the layer CF profile and the corresponding profile of in-cloud COD, the generator produces a set of 56 subcolumns by the following rules:

(1) Within each vertical layer the COD across cloudy subcolumns follows a beta PDF, whose variance profile is parameterized from the CF profile after Oreopoulos et al. (2012).

(2) Vertical overlap of both cloud occurrence and of the COD PDF ranks is controlled by the generalized overlap paradigm of Hogan and Illingworth (2000), in which the weighting factor between maximum and random overlap decays exponentially with the separation distance between two cloudy layers.

(3) The characteristic decorrelation length of cloud occurrence overlap, L_cf, and of the COD PDF rank correlation, L_cod, are treated as independent quantities, both parameterized as Gaussian functions of latitude and day of year. We use the updated Gaussian fits of O22b, which were derived from the same merged CloudSat-CALIPSO COD field that serves as our reference here; this is the strongest currently available form of the Räisänen generator on this dataset.

The generator is stochastic. To match the ML generator’s single-realization treatment, and the protocol under which SCOPS and the Räisänen generator were assessed in O22a, we run the generator once per scene with 56 subcolumns rather than averaging over many draws. This is acceptable because performance is ultimately assessed statistically over tens of thousands such scenes.

### 3.3 ML generator: two-stage CVAE-GAN + U-Net

The ML generator splits subcolumn generation into two stages that handle cloud occurrence (mask) and in-cloud optical depth separately. This matches the physical decomposition used in the Räisänen generator, where vertical overlap of cloud occurrence and the horizontal in-cloud COD PDF are controlled by independent parameterizations, while keeping each stage as a tractable supervised learning target.

A critical structural feature of our data processing and model evaluation pipeline is the sorting logic applied to the machine learning inputs and outputs. Unlike rudimentary approaches that sort subcolumns strictly layer-by-layer-which invariably destroys vertical coherence and skews the overlap statistics, our generator and the subsequent analysis pipeline rely on a column-integrated sorting logic. This method evaluates the integrated optical depth (or probability mass) over the entire atmospheric column to rank and order the subcolumns. By doing so, the generator strictly preserves the contiguous vertical structures and accurate decorrelation lengths necessary for proper radiative transfer integration.

**Stage 1 - Mask generator (CVAE-GAN)**

Stage 1 predicts the (39, 56) binary cloud mask given a 19-channel conditioning tensor assembled from the MERRA-2 large-scale state, scene scalars (with sin/cos encodings of Julian day and longitude), and layer values for cloud fraction, mean ice COD, and mean liquid COD.

The generator is a conditional variational autoencoder (CVAE). The posterior network q(z|x, m_obs) compresses the observed mask together with the conditioning tensor into a 64-dimensional Gaussian latent. The decoder p(m| x, z) is a U-Net with three downsampling and three upsampling stages built from residual blocks, operating on the (40, 56) scene plane after a one-layer is added as zero-pad to make the three stride-2 downsamples integer. The latent z is

projected to a 16-channel spatial map that is concatenated channelwise to the broadcast conditioning tensor before being fed into the U-Net. At training time z is parameterized anew from the posterior; at inference time z is drawn from the standard-normal prior.

A pure reconstruction loss produces overly contiguous vertical cloud patterns: the decoder prefers to minimize per-pixel error by extending cloud layers vertically, which tightens overlap beyond what is observed. To counteract this bias we add a PatchGAN-style discriminator (Isola et al., 2017) trained alongside the CVAE. The discriminator receives the mask together with the layer-broadcast conditioning tensor and outputs a $5 \times 7$ map of patchwise realism scores. Spectral normalization (Miyato et al., 2018) is applied to every discriminator convolution to stabilize training at high adversarial weights.

**Stage 2 - COD generator (U-Net)**

Stage 2 predicts the (39, 56, 2) field of log10(ice COD) and log10(liquid COD) given the Stage 1 mask and the same 19-channel conditioning tensor. The network is a deterministic U-Net with the same residual-block backbone as the Stage 1 decoder.

We use a deterministic architecture because the mask already encodes most of the subgrid stochasticity; conditional on a fixed mask, the COD is largely a function of the large-scale ice and liquid content and the meteorological state. To guarantee exact consistency with the large-scale ice and liquid layer-mean COD, the raw log-COD output of the U-Net is post-scaled layer-by-layer and phase-by-phase so that, averaged over the 56 subcolumns with the predicted mask as weight, the in-cloud mean COD equals exactly the large-scale input. This post-scaling is applied at both training and inference time. Stage 2 is trained independently of Stage 1, with the ground-truth mask as the mask input. This decoupling lets us evaluate errors in the mask and COD components separately.

**Training setup**

Both stages are trained with AdamW ($\beta = 0.5, 0.999$; weight decay $10^{-5}$), initial learning rate $3 \times 10^{-4}$, and ReduceLROnPlateau with factor 0.5 and patience 5 (Stage 1) or 7 (Stage 2). Stage 1 is trained for up to 150 epochs with early-stop patience 20; Stage 2 for up to 200 epochs with early-stop patience 25. Mixed-precision training uses bfloat16 autocast throughout. Training is distributed across four NVIDIA A100 GPUs on the NCCS DISCOVER cluster using PyTorch DistributedDataParallel. Gradient clipping at a maximum norm of 1.0 is applied to both generator and discriminator at every step; batches on which any rank reports a non-finite loss are skipped.

**Inference**

At inference the generator takes the same 19-channel conditioning tensor used during training and produces 56 subcolumns per scene: Stage 1 samples z from the prior and decodes to a probabilistic mask, which is binarized at 0.5; Stage 2 consumes that binary mask together with the conditioning tensor and outputs the post-scaled (log10 ice COD, log10 liquid COD) field, exponentiated back to linear units. The final per-scene output tensor is (39, 56, 3) with three channels (mask, ice COD, liquid COD) that match the format of the CloudSat reference and the Räisänen baseline.

**3.4 Offline radiative transfer**

To translate subcolumn fields into TOA radiation budgets, every scene is fed offline to the Rapid Radiative Transfer Model for GCMs (RRTMG; Iacono et al., 2008). For each scene we combine the 56-subcolumn field of (mask, $\tau_ice$, $\tau_liq$) with the matching MERRA-2 atmospheric column - temperature, water vapour, ozone, and pressure on the same 72-layer hybrid-sigma grid that conditions the generators, and the scene-mean (56-subcolumn average) ice and liquid effective radii derived from the MODIS-constrained CloudSat reference dataset.

Each subcolumn is integrated independently through RRTMG's 14 SW and 16 LW bands twice, once with the predicted cloud field and once cloud-free, yielding all-sky and clear-sky net fluxes at TOA and surface in order to produce CRE as the difference between the two. SW fluxes are converted to 24-h means using the daily-averaged solar weighting factor described in Section 2.2. Since the same RRTMG configuration, atmospheric profile, and microphysical parameterization are applied to all three subcolumn sources (REF, Räisänen, ML), any disagreement in TOA CRE is attributable to the cloud field alone.

## 4. Results

### 4.1 Sample subcolumn realizations

To set the stage with specific examples, Fig. 1 shows representative 56-subcolumn realizations on a single day (2007/07/01), one per row for three cloud regimes selected from the REF field: a tropical deep-convective scene (top, 152.4°E, 5.1°S), a tropical multilayer scene (middle, 178.0°E, 9.4°S), and a midlatitude stratocumulus deck (bottom, 154.5°W, 38.3°N, zoomed to 1000-680 hPa). Within each panel the subcolumns are reordered by descending column-integrated total $\tau$ so the three generators can be compared on the same axes regardless of their internal subcolumn ordering.

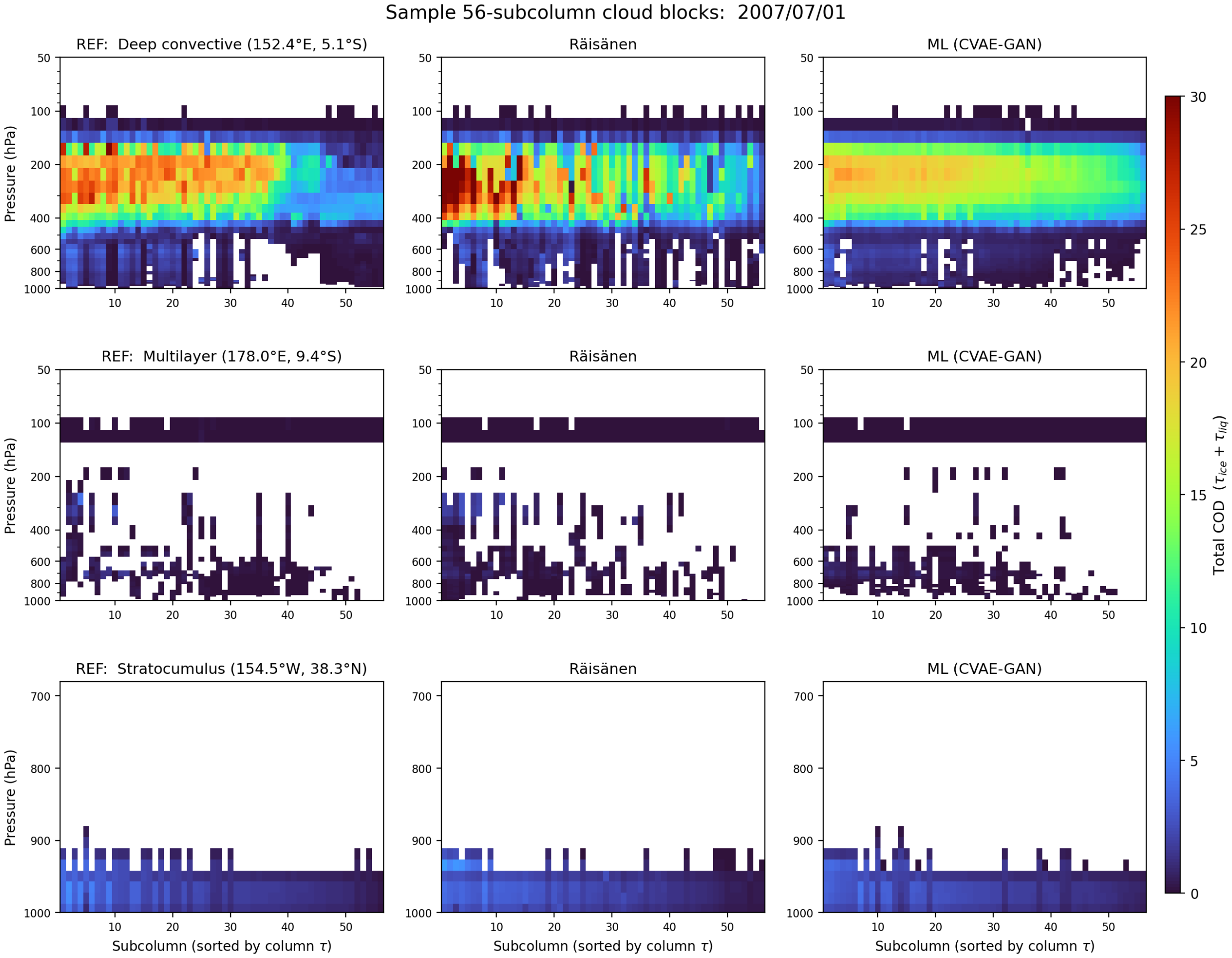

**Figure 1.** Representative 56-subcolumn cloud blocks for three cloud regimes from a single day (2007/07/01) of the held-out evaluation period: (top) deep convective at 152.4°E, 5.1°S; (middle) multilayer at 178.0°E, 9.4°S; (bottom) stratocumulus at 154.5°W, 38.3°N (zoomed to 1000-680 hPa). Columns show REF (CloudSat-CALIPSO of O22a), Räisänen with O22b decorrelation lengths, and ML (CVAE-GAN). Color corresponds to total in-cloud τ_ice + τ_liq; clear cells (TCA < 0.5) are white. Within each panel the 56 subcolumns are reordered by descending column-integrated total τ.

The three generators reproduce the regime-dependent vertical envelope of the REF field - a single optically thick deck spanning ~200-400 hPa in the deep-convective row, two vertically separated thin decks in the multilayer row, and a thin low-level deck in the stratocumulus row but the subcolumn structure differs visibly among the three. In the deep-convective example, the Räisänen subcolumns of the optically thicker rank quartiles are nearly horizontally homogeneous within each layer, with sharp $\tau$ contrasts across subcolumns rather than smooth gradients. This is a direct signature of the beta PDF of each layer combined with rank-correlated maximum overlap in a contiguous deck. The ML generator, in contrast, smooths the subcolumn transition and produces a $\tau$ envelope that decays continuously toward the low-rank columns, in qualitative agreement with the REF field. In the multilayer example all three generators correctly produce two decks, but Räisänen has a number of isolated cloudy cells filling the column between the two main decks, an apparent consequence of applying the exponential decorrelation profile at every $\Delta z$, while ML keeps the two decks more cleanly distinct. The stratocumulus example is the easiest case for all three generators and the three are visually similar. These qualitative impressions of the test cases are made quantitative by examining the statistics of the full datset in the subsections that follow.

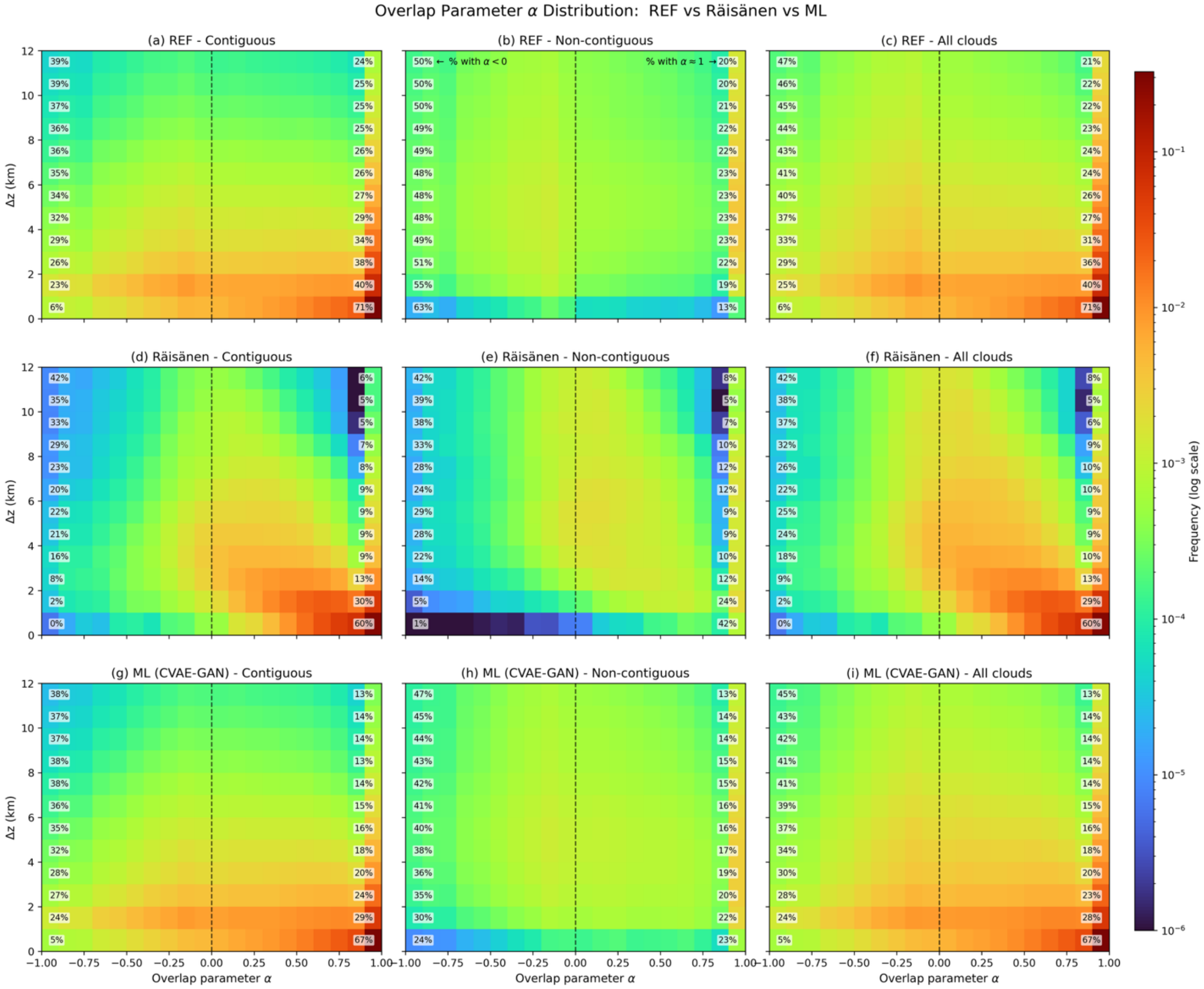

**Figure 2.** PDF of the cloud overlap parameter α = (C_true - C_rand) / (C_max - C_rand) as a function of inter-layer separation Δz, for (a-c) REF, (d-f) Räisänen, and (g-i) ML. Columns: contiguous layer pairs (a, d, g), non-contiguous pairs (b, e, h), and combined (c, f, i). α is computed for layer pairs in which both layers fall in the partly-cloudy regime (0.05 < CF_layer < 0.95). Edge annotations on each Δz row report the percentage of pairs with α < 0 (left) and α ∈ [0.9, 1.0] (right); legend on panel (b). The dashed line at α = 0 separates super-random (left) from sub-random (right) overlap. Color is on a log scale shared across panels.

### 4.2 Overlap statistics

Subcolumn fields are generated to serve as input to radiative transfer calculations, either offline or via McICA inside a host GCM (see Section 1). The two structural properties most directly responsible for the resulting fluxes are therefore the vertical overlap of the cloud mask and the rank ordering of in-cloud optical thickness across subcolumns. We diagnose the two separately using the overlap parameter $\alpha$ (Fig. 2) and the Spearman rank correlation $\rho$ (Fig. 3), as functions of inter-layer separation $\Delta z$, with a contiguous vs non-contiguous split.

The REF $\alpha$ distribution (Fig. 2a-c) has two physically distinct modes that the contiguous/non-contiguous split makes explicit. In the contiguous panel (a), where the two layers are joined by a column of cloudy intermediate layers, most of the population at small $\Delta z$ sits at $\alpha \approx 1$ (71% in the [0.9, 1.0] bin at $\Delta z$ = 0-1 km), the signature of a single coherent cloudy deck stacked layer-on-layer with maximum overlap. As $\Delta z$ grows, the $\alpha \approx 1$ mode weakens (24% at $\Delta z$ = 11-12 km) and a substantial $\alpha < 0$ super-random tail emerges (6% at $\Delta z$ = 0-1 km, 39% at $\Delta z$ = 11-12 km). In the non-contiguous panel (b), where the two layers are separated by at least one clear layer, the picture is reversed: the $\alpha < 0$ mode dominates at every $\Delta z$ (63% at $\Delta z$ = 0-1 km, 50% at $\Delta z$ = 11-12 km), the signature of two physically distinct, anti-correlated cloud decks coexisting in the same scene.

The Räisänen $\alpha$ distribution (Fig. 2d-f) succeeds where the underlying parameterization is built to succeed and fails where it is not. In the contiguous, small-$\Delta z$ regime, where exponential overlap reproduces the dominant maximum-overlap mode, Räisänen places 60% of pairs at $\alpha \approx 1$ (REF: 71%), a small underestimate. In the non-contiguous small-$\Delta z$ regime, however, Räisänen places only 1% of pairs at $\alpha < 0$ against REF's 63%, a near-total miss. The reason is structural: an exponential-decorrelation generator parameterized by a single $L_{cf}$ cannot produce the strong anti-correlation that two vertically separated cloud systems exhibit, because the formula cannot favor $\alpha < 0$ over $\alpha \geq 0$ at any specific $\Delta z$.

The ML generator (Fig. 2g-i) recovers both modes. In the contiguous panel ML has 67%/5% at small Δz (vs 71%/6% in REF) and 38%/13% at large Δz (vs 39%/24%). In the non-contiguous panel ML has 24% $\alpha < 0$ at small Δz - still an underestimate against REF's 63%, but a 24× improvement over Räisänen's 1%, and at large Δz ML's 47% closely matches REF's 50%. The ability of the ML generator to populate the super-random tail of α, especially in the non-contiguous regime where Räisänen is structurally precluded from doing so, is its single most consequential overlap improvement.

For the rank correlation ρ (Fig. 3), the contrast between the three generators is more subtle. At small Δz in contiguous pairs, Räisänen and REF both place 40% of pairs at $\rho \approx 1$ while ML overshoots slightly to 49%; both generators recover REF's small $\rho < 0$ tail to within a few percent (REF 7%, Räisänen 2%, ML 4%). At large Δz in contiguous pairs, however, REF develops a strong negative ρ tail (51% at $\Delta z$ = 11-12 km), an indication that the τ rank ordering is scrambled across thick cloud columns, and which Räisänen captures only partially (31%) while ML reproduces more closely (40%). In the non-contiguous panel the same pattern holds: at every Δz, ML's negative-ρ fraction sits between Räisänen's and REF's, closer to REF. Figs. 2 and 3 then provide an overall consistent picture of the Räisänen being well-calibrated where its analytical form aligns with the data and undershooting elsewhere, while ML tracks the REF distribution shape across the full Δz range.

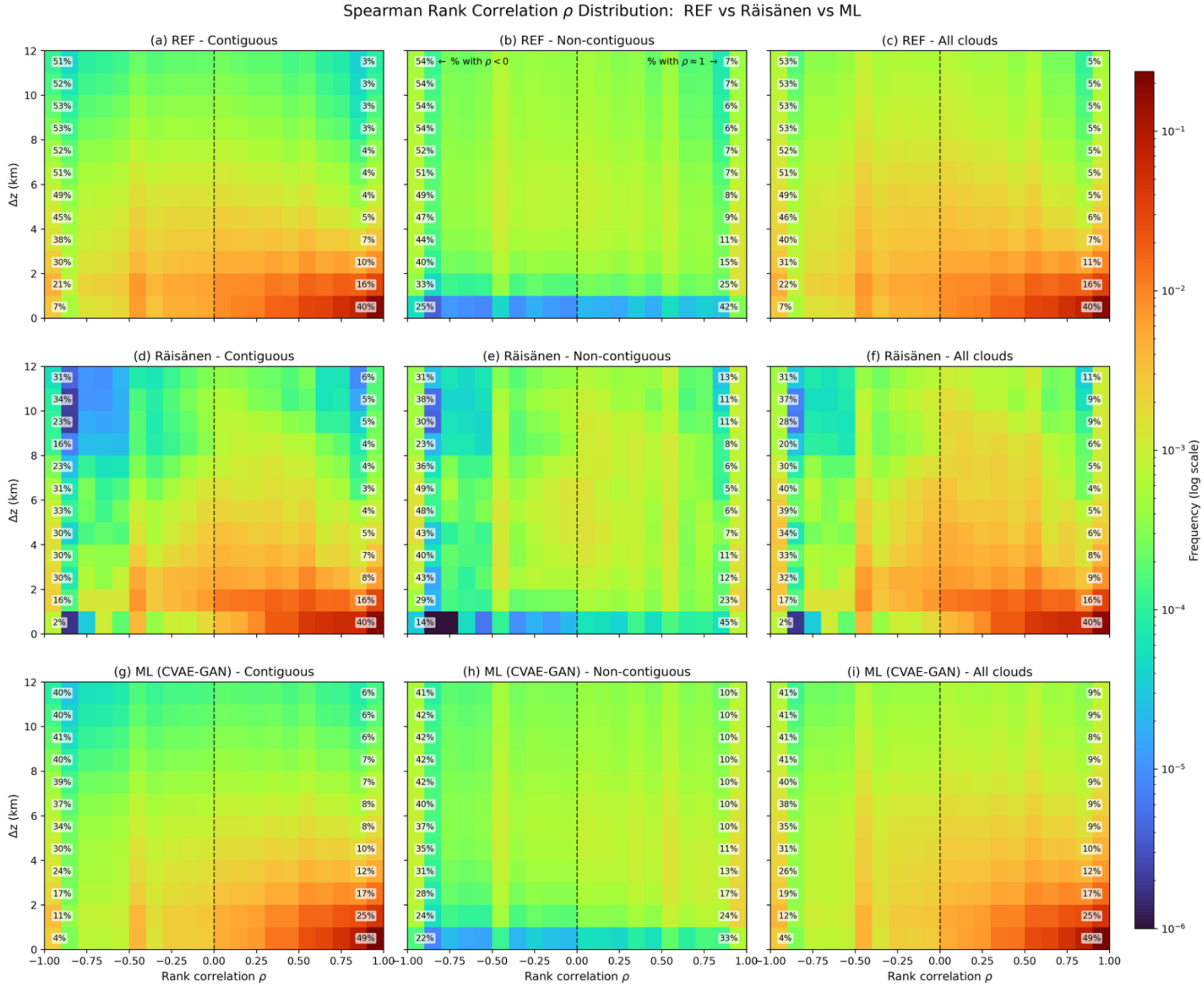

**Figure 3.** As Fig. 2 but for the in-cloud Spearman rank correlation ρ of τ_ice + τ_liq, computed across subcolumns that are cloudy at both layers. Panel layout, binning, and edge annotations follow Fig. 2.

### 4.3 Scene-level grid-cell statistics

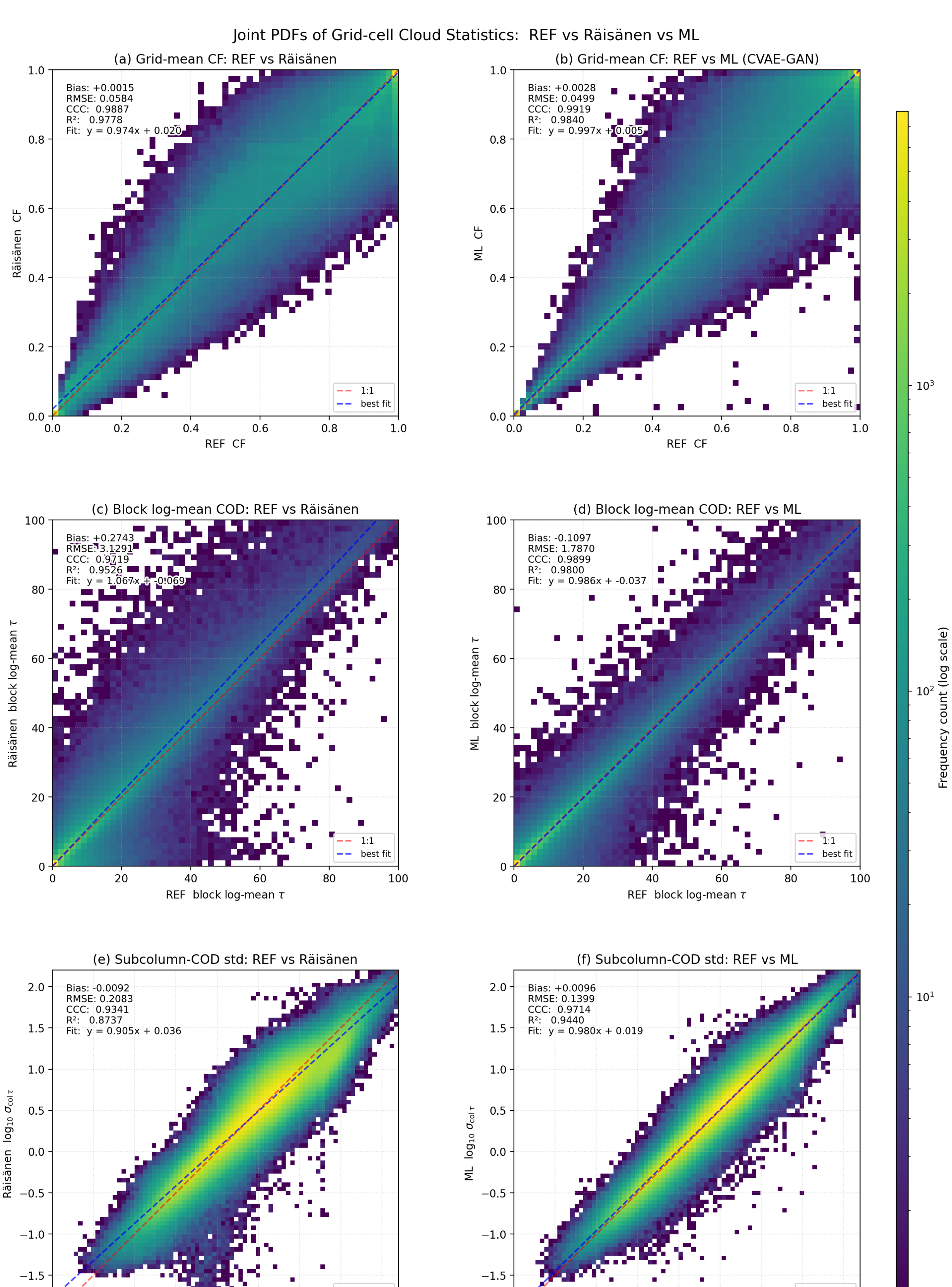

Joint PDFs of Grid-cell Cloud Statistics: REF vs Räisänen vs ML
(a) Grid-mean CF: REF vs Räisänen
Bias: +0.0015
RMSE: 0.0584
CCC: 0.9887
R²: 0.9778
Fit: y = 0.974x + 0.020
1:1
best fit
Räisänen CF
REF CF
(b) Grid-mean CF: REF vs ML (CVAE-GAN)
Bias: +0.0028
RMSE: 0.0499
CCC: 0.9919
R²: 0.9840
Fit: y = 0.997x + 0.005
ML CF
(c) Block log-mean COD: REF vs Räisänen
Bias: +0.2743
RMSE: 3.1291
CCC: 0.9719
R²: 0.9526
Fit: y = 1.067x + -0.069
Räisänen block log-mean τ
REF block log-mean τ
(d) Block log-mean COD: REF vs ML
Bias: -0.1097
RMSE: 1.7870
CCC: 0.9899
R²: 0.9800
Fit: y = 0.986x + -0.037
ML block log-mean τ
(e) Subcolumn-COD std: REF vs Räisänen
Bias: -0.0092
RMSE: 0.2083
CCC: 0.9341
R²: 0.8737
Fit: y = 0.905x + 0.036
Räisänen log10 σcol τ
REF log10 σcol τ
(f) Subcolumn-COD std: REF vs ML
Bias: +0.0096
RMSE: 0.1399
CCC: 0.9714
R²: 0.9440
Fit: y = 0.980x + 0.019
ML log10 σcol τ
Frequency count (log scale)

**Figure 4.** Joint PDFs of grid-cell summary statistics. (a, b) Grid-mean cloud fraction; (c, d) scene log-mean column $\tau$, $\exp(\langle \log(\tau_col + \varepsilon) \rangle)$; (e, f) log10 of the across-subcolumn standard deviation $\sigma_col$ $\tau$, with profiles below mean column-$\tau$ of 0.1 excluded. Left column: REF vs Räisänen; right column: REF vs ML. Each panel overlays the 1:1 line (red), OLS fit (blue), and a Bias/RMSE/CCC/$R^2$ fit summary. Counts on a log color scale, calibrated per panel.

The next test is whether the per-scene mean statistics are consistent between REF and each generator. Figure 4 shows joint PDFs of three such statistics: grid-mean cloud fraction (top), the block log-mean column optical thickness (middle), and the across-subcolumn standard deviation of $\tau_col$ in log10 (bottom).

For grid-mean cloud fraction (Fig. 4a, b), both generators are well-calibrated, but ML reduces the RMSE from 0.0584 (Räisänen) to 0.0499 and pushes the CCC from 0.989 to 0.992. The improvement is concentrated at high CF: visually, the off-1:1 populations above the diagonal in the Räisänen panel (scenes where Räisänen overestimates CF) are suppressed in the ML panel.

For the scene log-mean COD (Fig. 4c, d), the contrast is sharper. Räisänen has an RMSE of 3.13 with a fit slope of 1.07 and a low-end overestimate visible as a dense cloud above 1:1 at REF $\tau < 5$; ML reduces RMSE to 1.79 (a 43% reduction) with a fit slope of 0.99, and the off-diagonal populations collapse onto the 1:1 line. $R^2$ rises from 0.95 to 0.98. This is the regime where Räisänen's beta PDF with a single rank-correlation length tends to compress scene-mean COD toward the layer-mean; the ML generator is not constrained to that functional form and tracks REF more faithfully.

For the across-subcolumn $\sigma_col$ $\tau$ (Fig. 4e, f) ML again improves substantially. RMSE in log10 drops from 0.21 to 0.14, CCC from 0.93 to 0.97, $R^2$ from 0.87 to 0.94, and the slope tightens from 0.91 to 0.98. This metric matters because TOA shortwave reflectance is a convex function of $\tau_{col}$, so a generator that matches the secene-mean $\tau$ but underestimates the subcolumn spread will systematically overestimate TOA reflectance. In Section 4.5 we will see that the ML reduction in $\sigma_{col}$ $\tau$ RMSE yields smaller TOA SW biases.

### 4.4 ISCCP-style CTP-COT joint histogram

To connect with the protocol of O22a and O22b, under which all stochastic generators have been evaluated on this reference dataset, Fig. 5 shows the ISCCP-style CTP-COT joint histogram on the standard 7 × 6 grid. Panels (a-c) are the absolute JH for REF, Räisänen, and ML; panel (d) is Räisänen minus ML; panels (e, f) are the model minus REF biases. Each title reports the panel sum (cloudy fraction in %), the 42-cell RMS, and, for the bias panels, the scene-level Euclidean distance ED.

The Räisänen minus REF panel (Fig. 5e) reproduces the systematic biases identified in O22a and O22b: thin clouds are underestimated (e.g. -0.47 in the [180-310 hPa, $\tau \in$ 1.3-3.6] bin and -0.32 in the high-cloud thinnest column) while moderately optically thick mid-level clouds are overestimated (+0.27 at [180 hPa, $\tau \in$ 23-60]). The 42-cell RMS is 0.16 and the scene-level ED is 15.83. These biases are robust to the O22b retuning of the decorrelation lengths, which is consistent with O22b's own conclusion that the residual error is intrinsic to the analytical form of the Räisänen generator rather than to its specific coefficient values.

The ML minus REF panel (Fig. 5f) reduces almost every cell-level bias. The high-cloud thin underestimate is gone (-0.07 vs -0.47); the mid-level moderate-thick overestimate is reduced to within ±0.05; the 42-cell RMS drops from 0.16 to 0.08 and ED from 15.83 to 13.38. The largest remaining ML bias is +0.46 in the lowest-altitude/lowest-COT bin, which is shared with Räisänen (+0.41) and reflects a residual over-detection of the most optically thin low cloud common to both stochastic generators on this reference. The reduction of both the panel-mean RMS (a factor of two) and the scene-level ED demonstrates that the ML improvement is not a panel-averaging artifact but reflects scene-by-scene structural agreement with the reference.

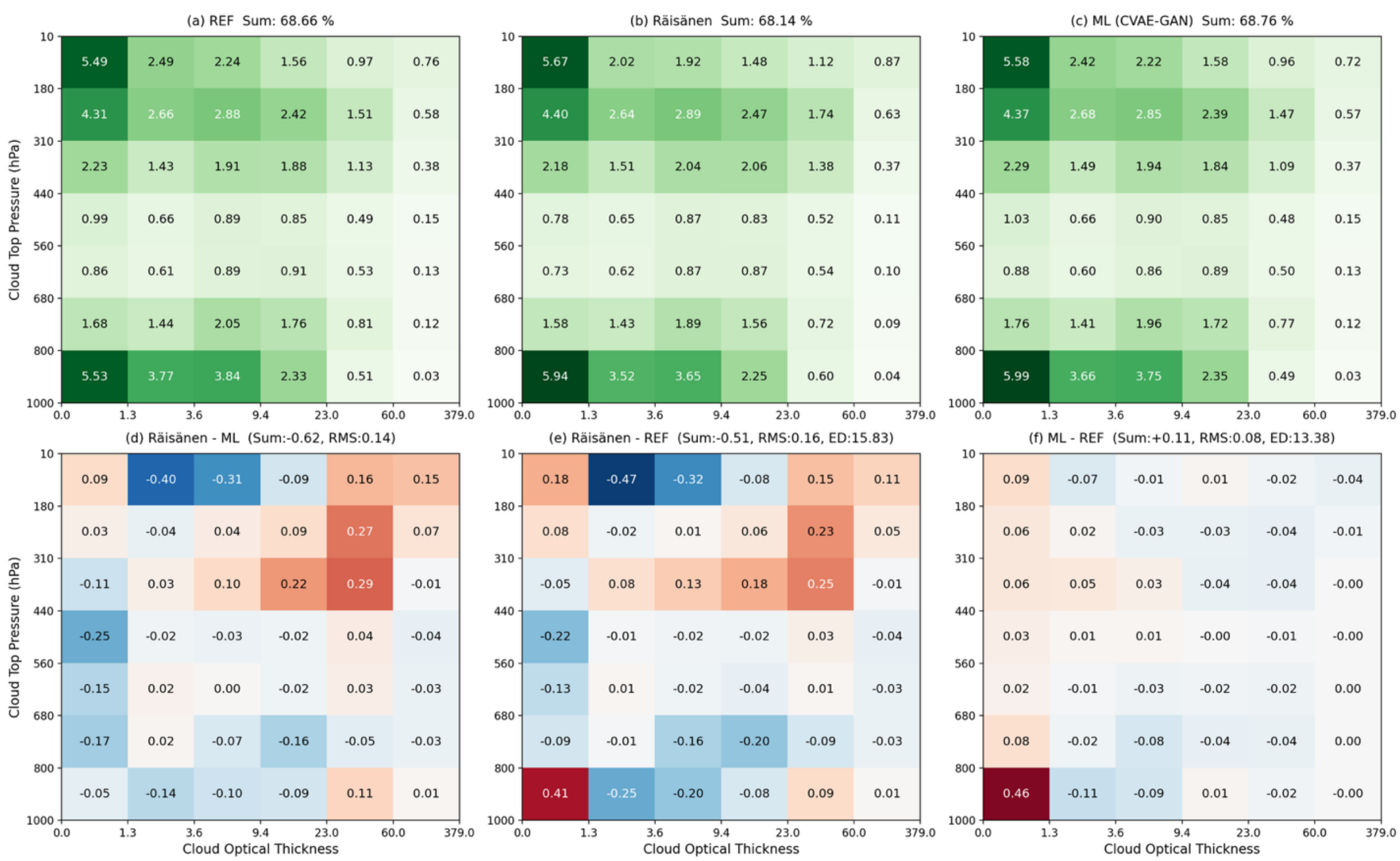

**Figure 5.** ISCCP-style CTP-COT joint histogram of the subcolumn cloud field on the standard 7 × 6 ISCCP-D grid. (a) REF, (b) Räisänen, (c) ML; (d) Räisänen minus ML; (e) Räisänen minus REF, (f) ML minus REF. Per-subcolumn COT is the column-integrated $\tau_ice + \tau_liq$; CTP is the highest level where cumulative top-down $\tau$ first exceeds $10^{-3}$. Counts are normalized to percent within each 10° latitude band and aggregated with cosine-latitude weights. Each title reports the panel sum (cloudy %), 42-cell RMS, and, for biases (e, f), the scene-level Euclidean distance ED computed as a mean from per-scene JHs.

### 4.5 Offline radiative transfer

To translate the JH improvements into the radiation budget itself, we run the offline RRTMG calculation described in Section 3.4. Figure 6 attributes the global-mean TOA CRE to (CTP, COT) cells (“cloud types”); Fig. 7 maps the resulting CF and TOA CRE biases globally on a 5° × 5° grid.

The REF panels of Fig. 6 (a, d, g) show the familiar pattern: optically thicker low clouds dominate the SW cooling (cell contributions reaching -3.6 W $m^{-2}$), high thin cirrus dominates the LW warming (+1.79 W $m^{-2}$ in the 10-180 hPa, $\tau \in$ 1.3-3.6 bin), and the Total CRE is negative everywhere except the high-thin column. The Räisänen minus REF color map (panels b, e, h) shows that Räisänen overestimates the SW cooling contribution from moderately thick mid-level cloud bins and underestimates the cirrus LW warming, with the two biases partially cancelling in Total but leaving a net Räisänen-vs-REF Total CRE bias of -0.69 W $m^{-2}$. The ML - REF color map (c, f, i) is visibly paler everywhere, with cell-level biases below ±0.1 W $m^{-2}$ across most cloud types; the global-mean Total CRE bias is -0.21 W $m^{-2}$ for ML against -0.69 W $m^{-2}$ for Räisänen, a factor-of-three reduction in absolute bias.

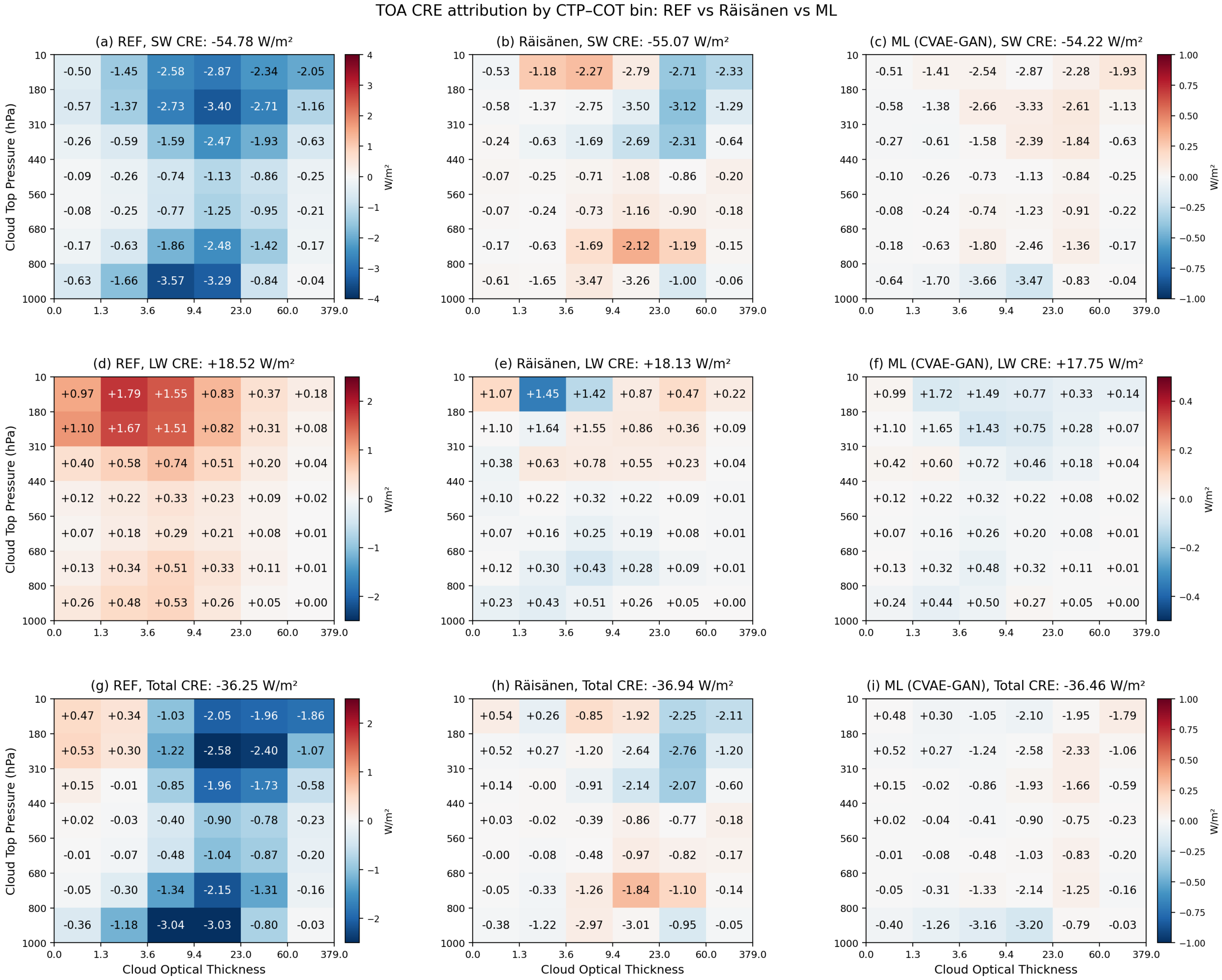

**Figure 6.** ISCCP-style attribution of the global-mean TOA CRE to (CTP, COT) cells ("cloud types"). Rows: SW (a-c), LW (d-f), Total CRE (g-i). Columns: REF, Räisänen, ML (CVAE-GAN). Each cell shows (mean cell CRE) × (cosine-latitude-weighted cell frequency); summing the 42 cells reproduces the panel-title global mean. CRE per subcolumn is computed offline from RRTMG fluxes (all-sky minus clear-sky net flux at TOA); SW CRE is converted to 24-h means via the standard daily solar-weighting factor. The REF column is colored by its own contribution; the Räisänen and ML columns retain their own contributions as printed numbers but are colored by the model minus REF bias.

Figure 7 maps these biases globally. The Räisänen CF bias map (d) shows the canonical structure documented by O22a and O22b - a pronounced negative bias along the ITCZ (deep blue, reaching -4%) and a positive bias over midlatitude oceans and arid land masses - that arises from the combination of the beta-PDF and the lat × DOY-fitted exponential overlap. The ML CF bias map (e) is essentially uniform within ±1%, with global RMSE of 0.49% against Räisänen's 1.80%. The SW TOA CRE bias map for Räisänen (i) inherits the same tropical pattern, with positive biases (less negative SW CRE than REF) over deep-convective regions reaching +6 W $m^{-2}$; the ML map (j) is again much flatter, with RMSE 1.23 vs 2.06 W $m^{-2}$ for Räisänen. The global-mean SW bias magnitudes are similar (Räisänen -0.31, ML +0.59 W $m^{-2}$), but the spatial RMSE difference is what matters for use in a coupled ESM, where regional rather than global radiation biases drive feedback errors. The LW maps (n, o) are the one place where Räisänen mean-beats ML. Although Räisänen's small global mean bias (-0.39 W $m^{-2}$) benefits from compensating positive and negative regional biases (Fig. 7n), it still achieves the smallest spatial RMSE of any LW model field (0.74 W $m^{-2}$),while ML carries a slightly larger global LW underestimate (-0.78 W $m^{-2}$) with comparable RMSE (0.94 W $m^{-2}$). This LW deficit traces to a small under-prediction of the highest-altitude thinnest clouds, visible as the negative tendency in the upper-left bins of Fig. 6f (which is still better than Fig. 6e overall), the most notable residual bias not addressed by the present generator.

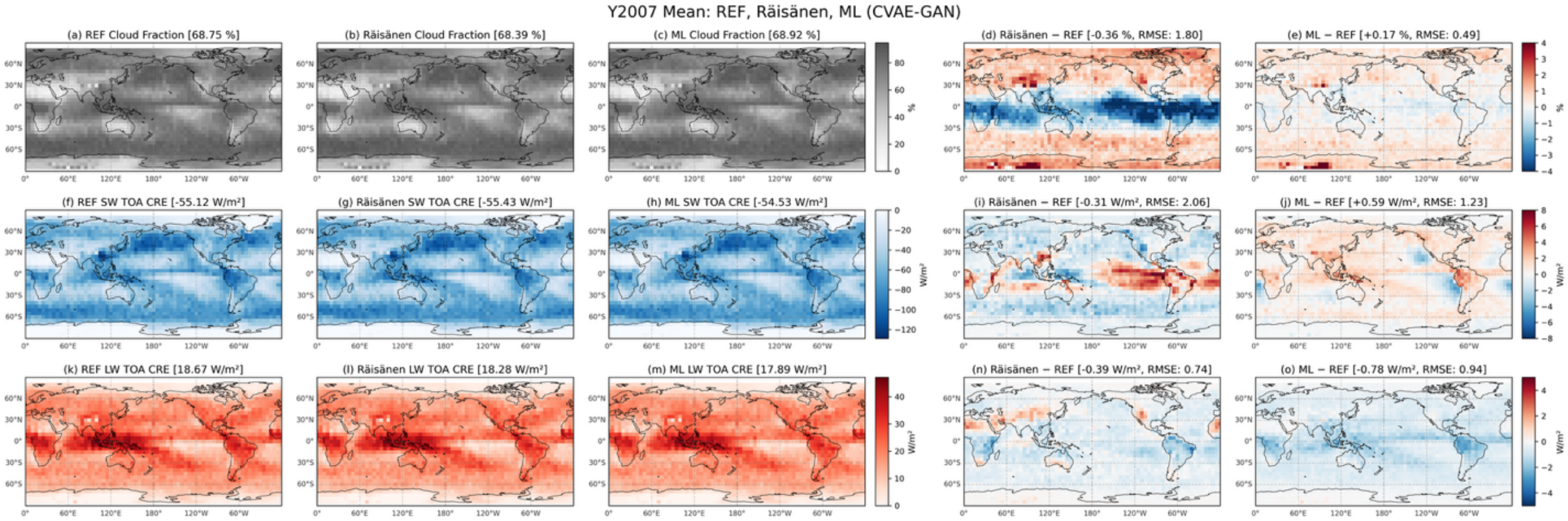

**Figure 7.** Year-2007 mean spatial distributions and biases. Rows: cloud fraction (a-e), SW TOA CRE (f-j), LW TOA CRE (k-o). Columns 1-3: absolute fields for REF, Räisänen, and ML on a 5° × 5° grid. Columns 4-5: model minus REF bias for Räisänen and ML, on row-symmetric colorbar limits (±4% for CF, ±8 W $m^{-2}$ for SW, ±5 W $m^{-2}$ for LW). Titles report cosine-latitude-weighted global means and, for biases, the spatial cosine-latitude-weighted RMSE. SW fluxes are 24-h means.

## 5. Summary and conclusions

We have developed a two-stage CVAE-GAN + U-Net machine-learning subcolumn generator for the GEOS atmospheric model and evaluated it against the merged CloudSat-CALIPSO COD field of Oreopoulos et al. (2022a) and the updated Räisänen baseline of Oreopoulos et al. (2022b) under which prior analytical generators have been benchmarked. The evaluation chain runs from low-level subcolumn structure to TOA radiation, and ML improves every link except one:

(i) **Overlap.** The ML generator captures REF's bimodal $\alpha(\Delta z)$ distribution including the super-random $\alpha < 0$ tail of non-contiguous layer pairs that an exponential-decorrelation parameterization is structurally unable to produce (Fig. 2). The optical-depth rank correlation $\rho$ matches the REF shape across $\Delta z$ more closely than Räisänen, particularly the negative $\rho$ tail at large $\Delta z$ (Fig. 3).

(ii) **Scene-mean statistics.** As a downstream consequence of better overlap and better in-cloud $\tau$ ordering, the per-scene mean cloud fraction, log-mean column $\tau$, and across-subcolumn $\sigma_col$ $\tau$ track REF more tightly than Räisänen does, with RMSE reductions of 15%, 43%, and 33% respectively and correlation coefficients exceeding 0.97 on all three (Fig. 4).

(iii) **ISCCP CTP-COT JH.** Under the standard O22a/O22b protocol, ML eliminates

the well-known Räisänen high-thin underestimate and mid-level moderate-thick overestimate, halving the panel-mean RMS (0.16 → 0.08) and tightening the scene-level Euclidean distance (15.83 → 13.38; Fig. 5).

(iv) **Offline TOA radiation.** The JH improvements translate directly into a smaller global-mean Total TOA CRE bias (0.21 W $m^{-2}$ for ML against 0.69 W $m^{-2}$ for Räisänen), a CF bias map RMSE that drops from 1.80% to 0.49%, and an SW TOA CRE bias map RMSE that drops from 2.06 to 1.23 W $m^{-2}$ (Figs. 6, 7). The single trade-off is a residual LW underestimate of -0.78 W $m^{-2}$ traceable to a slight under-prediction of the thinnest high cirrus. But even in the LW, the CRE attributed to cloud types matches the reference better than Räisänen, as with SW CRE.

The chain of evidence - better overlap → better block statistics → better JH → better radiation matches the chain of dependencies that a McICA-coupled ESM physics column traverses every time radiation is called. A version of the ML generator that can be run efficiently on CPUs will therefore be a candidate to serve as the default GEOS subcolumn generator for both interactive radiation and COSP, and will enable new assessments of GEOS cloud feedback per the methodology of Lee and Oreopoulos (2025). In short, online integration in GEOS with CPU efficiency, evaluation in an AMIP4K configuration, and a targeted fix of the residual high-cirrus underestimate are the natural next steps and are in progress.

**AI Disclosure Statement**

During the preparation of this work, the authors utilized generative AI tools, specifically Gemini (Google) and Claude (Anthropic), to assist in developing the machine learning codebase, writing data analysis and visualization scripts, and refining the manuscript's text. The authors carefully reviewed, verified, and tested all AI-generated code and text. The AI tools were not used to create, alter, or manipulate any original research data or results. The authors take full responsibility for the accuracy, originality, and integrity of the final published work.

**Figure captions**

**Figure 1.** Representative 56-subcolumn cloud blocks for three cloud regimes from a single day (2007/07/01) of the held-out evaluation period: (top) deep convective at 152.4°E, 5.1°S; (middle) multilayer at 178.0°E, 9.4°S; (bottom) stratocumulus at 154.5°W, 38.3°N (zoomed to 1000-680 hPa). Columns show REF (CloudSat-CALIPSO of O22a), Räisänen with O22b decorrelation lengths, and ML (CVAE-GAN). Color corresponds to total in-cloud $\tau_{ice} + \tau_{liq}$; clear cells (TCA < 0.5) are white. Within each panel the 56 subcolumns are reordered by descending column-integrated total $\tau$.

**Figure 2.** PDF of the cloud overlap parameter $\alpha = (C_{true} - C_{rand}) / (C_{max} - C_{rand})$ as a function of inter-layer separation $\Delta z$, for (a-c) REF, (d-f) Räisänen, and (g-i) ML. Columns: contiguous layer pairs (a, d, g), non-contiguous pairs (b, e, h), and combined (c, f, i). $\alpha$ is computed for layer pairs in which both layers fall in the partly-cloudy regime ($0.05 < CF_{layer} < 0.95$). Edge annotations on each $\Delta z$ row report the percentage of pairs with $\alpha < 0$ (left) and $\alpha \in [0.9, 1.0]$ (right); legend on panel (b). The dashed line at $\alpha = 0$ separates super-random (left) from sub-random (right) overlap. Color is on a log scale shared across panels.

**Figure 3.** As Fig. 2 but for the in-cloud Spearman rank correlation $\rho$ of $\tau_{ice} + \tau_{liq}$, computed across subcolumns that are cloudy at both layers. Panel layout, binning, and edge annotations follow Fig. 2.

**Figure 4.** Joint PDFs of grid-cell summary statistics. (a, b) Grid-mean cloud fraction; (c, d) scene log-mean column $\tau$, $\exp(\langle \log(\tau_{col} + \varepsilon) \rangle)$; (e, f) log10 of the across-subcolumn standard deviation $\sigma_{col}$ $\tau$, with profiles below mean column-$\tau$ of 0.1 excluded. Left column: REF vs

Räisänen; right column: REF vs ML. Each panel overlays the 1:1 line (red), OLS fit (blue), and a Bias/RMSE/CCC/$R^2$ fit summary. Counts on a log color scale, calibrated per panel.

**Figure 5.** ISCCP-style CTP-COT joint histogram of the subcolumn cloud field on the standard 7 × 6 ISCCP-D grid. (a) REF, (b) Räisänen, (c) ML; (d) Räisänen minus ML; (e) Räisänen minus REF, (f) ML minus REF. Per-subcolumn COT is the column-integrated $\tau_ice$ + $\tau_liq$; CTP is the highest level where cumulative top-down $\tau$ first exceeds $10^{-3}$. Counts are normalized to percent within each 10° latitude band and aggregated with cosine-latitude weights. Each title reports the panel sum (cloudy %), 42-cell RMS, and, for biases (e, f), the scene-level Euclidean distance ED computed as a mean from per-scene JHs.

**Figure 6.** ISCCP-style attribution of the global-mean TOA CRE to (CTP, COT) cells ("cloud types"). Rows: SW (a-c), LW (d-f), Total CRE (g-i). Columns: REF, Räisänen, ML (CVAE-GAN). Each cell shows (mean cell CRE) × (cosine-latitude-weighted cell frequency); summing the 42 cells reproduces the panel-title global mean. CRE per subcolumn is computed offline from RRTMG fluxes (all-sky minus clear-sky net flux at TOA); SW CRE is converted to 24-h means via the standard daily solar-weighting factor. The REF column is colored by its own contribution; the Räisänen and ML columns retain their own contributions as printed numbers but are colored by the model minus REF bias.

**Figure 7.** Year-2007 mean spatial distributions and biases. Rows: cloud fraction (a-e), SW TOA CRE (f-j), LW TOA CRE (k-o). Columns 1-3: absolute fields for REF, Räisänen, and ML on a 5° × 5° grid. Columns 4-5: model minus REF bias for Räisänen and ML, on row-symmetric colorbar limits (±4% for CF, ±8 W $m^{-2}$ for SW, ±5 W $m^{-2}$ for LW). Titles report cosine-latitude-weighted global means and, for biases, the spatial cosine-latitude-weighted RMSE. SW fluxes are 24-h means